# Comparative First-Principles Study of the Physical Properties, Phonon-Mediated Superconductivity and Functional Potential of Technetium-Based Binary Borides $TcB_2$, $Tc_2B$ and $Tc_3B$

M. Rabi Us Sani Shapon, Jubair Hossan Abir, Md. Asif Afzal, Raihana Shams Islam*, Saleh Hasan Naqib*

[1]Department of Physics, University of Rajshahi, Rajshahi 6205, Bangladesh

*Corresponding authors; Emails: ***jsr_eda1@ru.ac.bd** (Saleh Hasan Naqib), *__rsislam@ru.ac.bd__ (Raihana Shams Islam)

## Abstract

Phonon-mediated superconductivity in binary transition-metal borides reflects a balance between the metal *d*-states, metal-boron bonding, and the lattice vibrations that couple to them. Here, we compare *P6/mmm* $TcB_2$ at 20 GPa with *I4/mcm* $Tc_2B$ and *P4/mmm* $Tc_3B$ at 0 GPa within a first-principles framework. Their mechanical responses differ appreciably: $TcB_2$ is the most rigid under its compressed condition, $Tc_2B$ shows the strongest tendency toward ductility, and $Tc_3B$ falls between the two. Despite these differences, all three phases remain metallic. The states at the Fermi level are dominated by Tc-4*d* character, although the extent of Tc-*d*/B-*p* mixing varies from one compound to another. Charge redistribution is more directional and B-centered in $TcB_2$, but more delocalized around the Tc network in $Tc_2B$ and $Tc_3B$. Inclusion of spin-orbit coupling leaves all three compounds metallic. The optical response remains metallic and electric field polarization dependent, with substantial visible reflectivity and strong ultraviolet absorption. Low-frequency Tc rich vibrations dominate the electron-phonon coupling (EPC). With $\mu^* = 0.10$, the $T_c$ values of $TcB_2$, $Tc_2B$, and $Tc_3B$ are 24.92, 11.07, and 15.12 K, respectively. Although $Tc_3B$ exhibits a higher electronic density of states at the Fermi level, $N(E_F)$, $TcB_2$ couples most strongly to EPC-active low-frequency modes, showing that the superconducting hierarchy depends on electronic-phonon matching rather than DOS magnitude alone.

## 1 Introduction

In conventional superconductors, lattice vibrations mediate an effective attraction between electrons, enabling Cooper pairing and the superconducting state below a critical temperature **[1]**. First-principles electron-phonon theory now makes it possible to relate this pairing interaction to the electronic states and lattice vibrations that contribute to it **[2]**. The same mechanism can produce very high transition temperatures in hydrogen-rich systems: sulfur hydride exceeds 200 K under strong compression **[3]**, while lanthanum superhydrides reach still higher values under high pressures **[4,5]**. Such extreme compression, however, motivates the search for phonon-mediated superconductors with much milder pressure requirements. $MgB_2$ remains the benchmark among boride superconductors, with a transition temperature of 39 K at ambient pressure **[6]**, and its superconductivity is associated with B-derived $\sigma$-states strongly coupled to in-plane B bond-stretching modes **[7–11]**. More broadly, light B atoms can support high-frequency phonons, while changes in metal-B bonding and electronic structure can reshape the states near $E_F$ **[7–11]**. In binary borides, the superconducting response is therefore governed by the interplay between the electronic states present at $E_F$ and the phonon modes to which those electronic states couple **[7–11]**.

The contrast is especially clear when transition-metal *d*-states contribute directly near $E_F$. $FeB_4$, for example, was predicted and then realized as a superconducting hard boride, showing that strong bonding networks and conventional superconductivity can coexist in transition-metal borides **[12,13]**. First-principles and data-driven surveys of $MgB_2$-related structures further show that changing the metal species can reorganize both the Fermi-level electronic states and the relevant phonon spectrum **[14,15]**. $MoB_2$ provides a clearer departure from the $MgB_2$ mechanism. Calculations under pressure identify substantial Mo-*d* character at $E_F$ and important coupling to lower-frequency metal vibrations **[16]**, consistent with the observed pressure-induced superconducting phase **[17]**. Mode-resolved calculations across several transition-metal diborides likewise show that *d*-electron coupling to metal-dominated acoustic modes can become the main superconducting channel rather than the B-derived optical-mode coupling characteristic of $MgB_2$ **[18]**. Changes in crystal symmetry and metal-B hybridization reshape the states available for pairing, and the bonding network changes the character and stiffness of the phonons that couple to them **[16–18]**. For transition-metal borides, the relevant question is how changes in crystal structure and bonding reorganize both the *d*-derived electronic states and the phonons that couple to them.

The Tc-B system brings this problem into a set of closely related compounds with different stoichiometries and crystal symmetries. Experimental work established technetium boride phases decades ago **[19]**. First-principles studies later examined $TcB_2$ and its physical properties **[20]**. Broader phase searches then expanded the structural picture of Tc-B under pressure **[21–23]**, with separate studies extending it to $TcB_3$ and $TcB_4$ **[24–26]**. The literature therefore describes a structurally diverse Tc–B chemistry with substantial attention to mechanical response **[27]**, but superconductivity was addressed only recently **[28]**. Tao et al. reported superconducting technetium-boride candidates at moderate or ambient pressure, including *P6/mmm* $TcB_2$ at 20 GPa and *I4/mcm* $Tc_2B$ and *P4/mmm* $Tc_3B$ at 0 GPa **[28]**. Their electronic and electron–phonon calculations attributed the dominant coupling in these phases to Tc-4*d* states near $E_F$ interacting mainly with low-frequency Tc vibrations **[28]**, consistent with the *d*-electron/metal-phonon behavior reported for other transition-metal borides **[16,18]**. Despite this progress, the literature does not provide a unified comparison that connects structural and mechanical behavior with the superconducting electronic-phonon response across $TcB_2$, $Tc_2B$, and $Tc_3B$ **[20,22,27,28]**. As a result, it remains unclear how the different symmetry and bonding environments of $TcB_2$, $Tc_2B$, and $Tc_3B$ reorganize their electronic structure and lattice dynamics, and how those changes are reflected across their wider physical response. A direct comparison of these phases is needed to determine which features are common to a Tc-derived electronic framework and which depend sensitively on local bonding topology and crystal symmetry.

The present study treats $TcB_2$, $Tc_2B$, and $Tc_3B$ as a comparative set within one first-principles framework. Structural and mechanical responses are examined together with the electronic and bonding picture, and these are related to direction-dependent optical behavior and to the lattice dynamics governing electron-phonon coupling and phonon-mediated superconductivity. The calculations focus on *P6/mmm* $TcB_2$ at 20 GPa, *I4/mcm* $Tc_2B$ at 0 GPa, and *P4/mmm* $Tc_3B$ at 0 GPa under the conditions considered in this work. Because $TcB_2$ is evaluated under pressure whereas $Tc_2B$ and $Tc_3B$ are considered at ambient pressure, differences among the three phases are interpreted as the combined outcome of composition, symmetry, bonding environment, and pressure rather than as purely stoichiometric trends. We use the comparison to distinguish the Tc-derived electronic features shared across the series from those that are altered by structural changes strongly enough to modify lattice coupling and superconducting response. The structure-property trends across the three phases provide screening cues for related transition-metal borides, particularly for analogues that retain favorable *d*-state, bonding, and phonon characteristics.

## 2 Computational Methodology

The literature-reported $TcB_2$, $Tc_2B$, and $Tc_3B$ phases **[28]** were investigated at 20 GPa, 0 GPa, and 0 GPa, respectively. CASTEP **[29]** calculations employed PBE-GGA **[30]** for structural relaxation, finite-strain elastic calculations, electron-density-difference analysis, and electric field polarization-dependent optical properties. Plane-wave cutoff/k-point settings were 500 eV/15×15×12 for $TcB_2$, 650 eV/14×14×16 for $Tc_2B$, and 500 eV/18×18×8 for $Tc_3B$, with 0.10 eV Gaussian smearing. Mechanical stability was assessed using the appropriate zero-pressure elastic criteria for the tetragonal phases **[31]** and the stress-corrected formulation for $TcB_2$ at 20 GPa **[32]**. Polycrystalline elastic moduli and anisotropy were obtained from the elastic tensor using standard averaging and anisotropy formalisms **[33,34]**, while elastic-derived acoustic, Debye, and lattice-thermal descriptors followed established approaches **[35,36]**. Optical spectra were evaluated for [100] and [001] electric field orientations with a 3.0 eV plasma energy, 0.05 eV Drude broadening, and 0.5 eV energy smearing.

Quantum ESPRESSO (QE) **[37,38]** was used for the complementary electronic and lattice-dynamical workflow. Each phase was QE-relaxed at 20 GPa for $TcB_2$ and 0 GPa for $Tc_2B$ and $Tc_3B$ using PBEsol **[39]**, the corresponding PAW pseudopotentials, and 80/640 Ry wave-function/charge-density cutoffs. These structures were then used for non-SOC/SOC band structure calculations and the finalized non-SOC Fermi surfaces, visualized with FermiSurfer **[40]**. Phonon dispersion and EPC calculations were calculated using density-functional perturbation theory (DFPT) **[41]**. The interactions between the atomic nuclei and valence electrons were described using optimized norm-conserving Vanderbilt pseudopotentials from PseudoDojo library **[42,43]**. The wave-function/charge-density cutoffs were 80/640 Ry and 24×24×24 k-point meshes, 3×3×3 q-point mesh for EPC calculations. Metallic occupations were treated with Marzari-Vanderbilt smearing using degauss = 0.05 Ry **[44]**. The phonon/EPC calculations were carried out without SOC. The superconducting transition temperature was estimated using the isotropic Allen-Dynes-modified McMillan expression **[45,46]** with a repulsive Coulomb pseudopotential of $\mu^* = 0.10$.

## 3 Results and discussion

### 3.1 Structural Properties

The relaxed structures of $TcB_2$, $Tc_2B$, and $Tc_3B$ preserve the *P6/mmm*, *I4/mcm*, and *P4/mmm* symmetries, respectively, under the pressure conditions adopted here (20 GPa for $TcB_2$ and 0 GPa for $Tc_2B$ and $Tc_3B$). These are the lowest pressures at which Tao et al. reported dynamical stability for the corresponding superconducting phases **[28]**. Hexagonal $TcB_2$ (space group No. 191) contains Tc at the 1a and B at the 2d sites, giving alternating Tc planes and planar B honeycomb sheets. In tetragonal $Tc_2B$ (No. 140), Tc occupies the 8h and B the 4a sites, whereas $Tc_3B$ (No. 123) contains nonequivalent Tc sites at 2h and 1a together with B at 1b **[28]**. These contrasting stacking patterns, coordination environments, and Tc:B ratios provide the structural basis for comparing the mechanical, electronic, optical, and lattice-dynamical responses of the three phases.

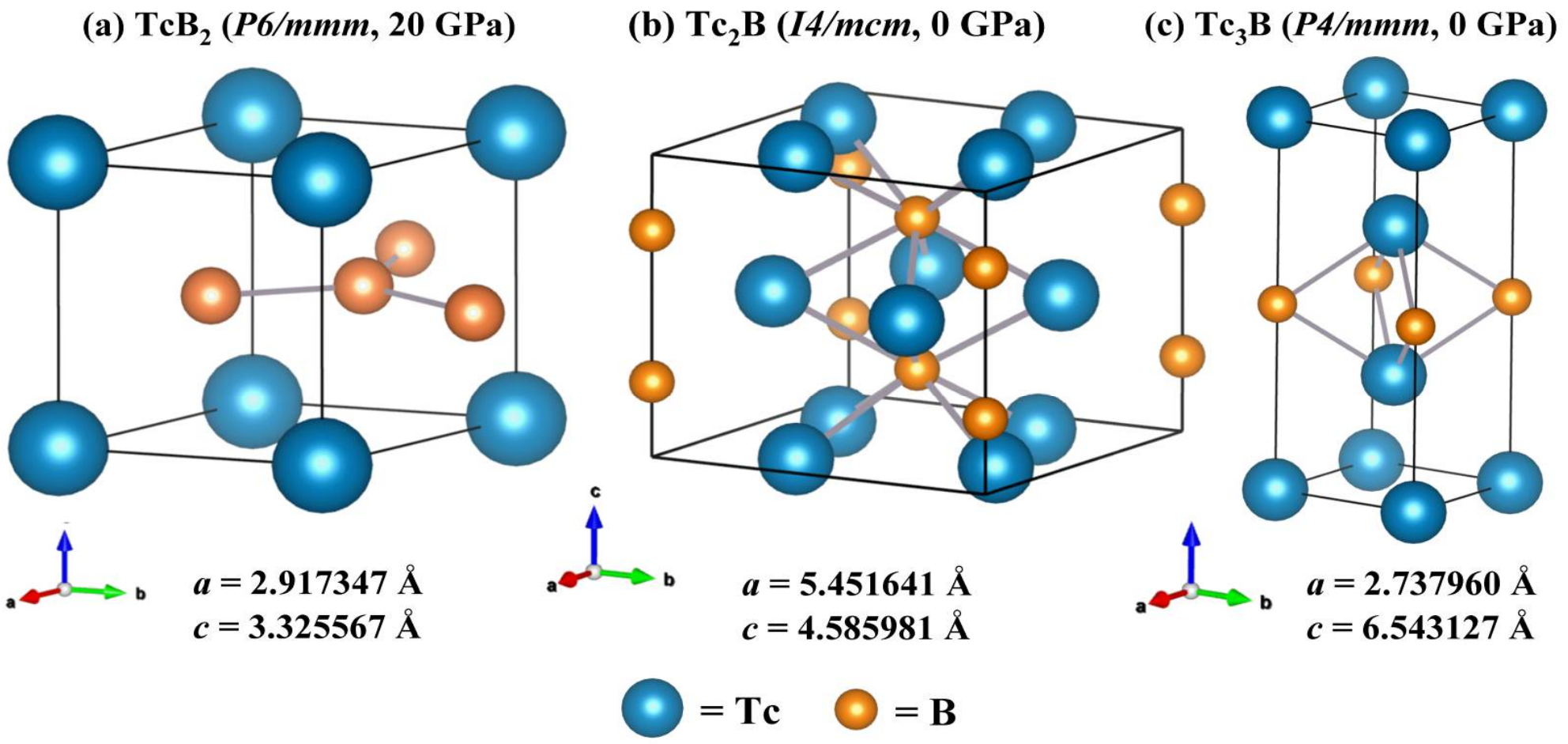

**Figure 1.** Optimized crystal structures of (a) $TcB_2$ (*P6/mmm*) at 20 GPa, (b) $Tc_2B$ (*I4/mcm*) at 0 GPa, and (c) $Tc_3B$ (*P4/mmm*) at 0 GPa. Blue and orange spheres denote Tc and B atoms, respectively.

The optimized lattice parameters, cell volumes, and mass densities are listed in **Table 1**. For the hexagonal cell, the volume is obtained from:

$$V_{hx} = \left(\frac{\sqrt{3}}{2}\right) a^2 c, \qquad (1)$$

where $a$ and $c$ are the basal-plane and axial lattice parameters, respectively. For the tetragonal cells, the corresponding relation is:

$$V_{tet} = a^2 c, \qquad (2)$$

The crystal density can be expressed as:

$$\rho = \frac{ZM}{N_A V}, \tag{3}$$

where $Z$ is the number of formula units in the conventional cell, $M$ is the molar mass of one formula unit, $N_A$ is Avogadro's constant, and $V$ is the unit-cell volume. The optimized densities are 8.165, 10.167, and 10.411 g cm$^{-3}$ for $TcB_2$, $Tc_2B$, and $Tc_3B$, respectively. The higher densities of the Tc-rich phases reflect the larger Tc fraction and the compact optimized cells.

**Table 1.** Comparison of the optimized structural parameters of $TcB_2$, $Tc_2B$, and $Tc_3B$ with phase- and pressure-matched values reported by Tao et al. **[28]**.

| Phase | *a* (Å) | *c* (Å) | *c/a* | *V* (Å$^3$) | ρ (g cm$^{-3}$) | Reference |
|---|---|---|---|---|---|---|
| $TcB_2$ | 2.917347 | 3.325567 | 1.1399 | 24.511648 | 8.165 | This work |
| | 2.90764 | 3.33220 | 1.1460 | — | — | Tao et al. **[28]** |
| $Tc_2B$ | 5.451641 | 4.585981 | 0.8412 | 136.297117 | 10.167 | This work |
| | 5.45112 | 4.58875 | 0.8418 | — | — | Tao et al. **[28]** |
| $Tc_3B$ | 2.737960 | 6.543127 | 2.3898 | 49.050042 | 10.411 | This work |
| | 2.73625 | 6.53966 | 2.3900 | — | — | Tao et al. **[28]** |

Agreement with the parent study was quantified using the relative deviation given by:

$$\delta_x = \left[\frac{(x_{this\ work} - x_{ref})}{x_{ref}}\right] \times 100, \tag{4}$$

where x represents either a or c. The deviations in a and c are +0.334% and −0.199% for $TcB_2$, +0.0096% and −0.0603% for $Tc_2B$, and +0.0625% and +0.0530% for $Tc_3B$, respectively. All optimized lattice constants therefore differ by less than 0.5% from the phase- and pressure-matched values of Tao et al. **[28]**, providing a direct validation of the structural models used here. Earlier first-principles studies of technetium diborides likewise show that equilibrium lattice metrics are sensitive to polymorph and pressure **[27,47]**, which is why unmatched phases are not mixed into the numerical benchmark. Viewed together with the contrasting cell geometries and layer topologies, this agreement provides a controlled structural basis for examining how stoichiometry and crystal symmetry reshape the elastic, electronic, optical, phonon, and electron–phonon responses.

## 3.2 Elastic, Mechanical and Thermophysical Properties

The elastic tensors of $TcB_2$, $Tc_2B$ and $Tc_3B$ were examined to establish pressure-appropriate mechanical stability and to compare their deformation resistance, empirical ductility, model hardness and elastic anisotropy. $TcB_2$ is considered at 20 GPa, whereas $Tc_2B$ and $Tc_3B$ are evaluated at 0 GPa. Consequently, comparisons involving $TcB_2$ reflect both composition and pressure. Zero-stress criteria were adopted for the tetragonal phases [31], while the hydrostatic stress correction of Barron and Klein was applied to $TcB_2$ [32].

### 3.2.1 Elastic tensor and mechanical stability

For hydrostatic compression P, the effective elastic tensor can be expressed in the stress-corrected Barron-Klein form [32]:

$$B_{ijkl} = C_{ijkl} - \mathrm{P}(\delta_{ik}\delta_{jl} + \delta_{il}\delta_{jk} - \delta_{ij}\delta_{kl}) \tag{5}$$

Here $\delta_{ij}$ denotes the Kronecker delta. The relevant pressure-corrected coefficients are:

$$\begin{aligned} B_{11} &= C_{11} - \mathrm{P}, & B_{33} &= C_{33} - \mathrm{P} \\ B_{44} &= C_{44} - \mathrm{P}, & B_{66} &= C_{66} - \mathrm{P} \\ B_{12} &= C_{12} + \mathrm{P}, & B_{13} &= C_{13} + \mathrm{P} \end{aligned} \tag{6}$$

Mechanical stability at finite pressure must therefore be assessed using $B_{ij}$ rather than the uncorrected $C_{ij}$ in **Table 2**. For hexagonal symmetry, the pressure-appropriate stability conditions [32] are:

$$\begin{aligned} B_{11} > |B_{12}|, \quad B_{44} > 0, \\ (B_{11} + B_{12})\, B_{33} > 2B_{13}^2 \end{aligned} \tag{7}$$

with $B_{66} = (B_{11} - B_{12})/2$. At zero pressure, $B_{ij} = C_{ij}$. The tetragonal phases additionally satisfy $C_{11} > |C_{12}|$, $C_{44} > 0$, $C_{66} > 0$ and $(C_{11} + C_{12})C_{33} > 2C_{13}^2$, following Mouhat and Coudert [31].

At 20 GPa, $TcB_2$ has stress-corrected coefficients $B_{11} = 654.31$, $B_{12} = 238.67$, $B_{13} = 243.78$, $B_{33} = 583.36$, $B_{44} = 82.65$ and $B_{66} = 207.82$ GPa. These coefficients satisfy the required positive-definiteness conditions. The reported elastic tensors of $Tc_2B$ and $Tc_3B$ also satisfy their respective zero-pressure criteria.

All three compounds exhibit $C_{11} > C_{33}$ (**Table 2**), indicating greater longitudinal stiffness within the basal plane than along c. $TcB_2$ has the largest longitudinal elastic coefficients at 20 GPa, while differing $C_{44}$ and $C_{66}$ values reveal distinct shear responses. Tao et al. [28] report data for the corresponding compositions and pressure conditions, enabling a useful computational comparison subject to the $Tc_2B$ phase-label discrepancy noted in

**Table 2**. The present values agree more closely with the reported $TcB_2$ values than with those for the tetragonal compositions [28]. Differences in the shear coefficients are also reflected in the calculated shear moduli, model hardness and elastic anisotropy.

**Table 2.** Calculated single-crystal elastic constants of $TcB_2$, $Tc_2B$, and $Tc_3B$ compared with the same-phase values reported by Tao et al. [28]. All elastic constants are in GPa.

| Compound | $P$ (GPa) | Space group | $C_{11}$ | $C_{12}$ | $C_{13}$ | $C_{33}$ | $C_{44}$ | $C_{66}$ | Reference |
|---|---|---|---|---|---|---|---|---|---|
| $TcB_2$ | 20 | *P6/mmm* | 674.31 | 218.67 | 223.78 | 603.36 | 102.65 | 227.82 | This work |
| | | | 685 | 213 | 220 | 589 | 85 | 236 | [28] |
| $Tc_2B$ | 0 | *I4/mcm* | 510.40 | 218.94 | 269.96 | 419.59 | 137.22 | 166.28 | This work |
| | | | 547 | 171 | 227 | 526 | 123 | 164 | [28] |
| $Tc_3B$ | 0 | *P4/mmm* | 551.86 | 149.89 | 249.00 | 436.68 | 169.39 | 116.76 | This work |
| | | | 449 | 229 | 241 | 490 | 128 | 202 | [28] |

### 3.2.2 Polycrystalline elastic moduli and deformation response

For polycrystalline aggregates, the Voigt and Reuss estimates provide upper and lower elastic bounds, whose arithmetic means define the Hill moduli [33]:

$$B_{\mathrm{H}} = \frac{B_{\mathrm{V}} + B_{\mathrm{R}}}{2}, \qquad G_{\mathrm{H}} = \frac{G_{\mathrm{V}} + G_{\mathrm{R}}}{2} \tag{8}$$

Young's modulus and Poisson's ratio were then obtained from the Hill bulk and shear moduli [33]:

$$E = \frac{9B_{\mathrm{H}}G_{\mathrm{H}}}{3B_{\mathrm{H}} + G_{\mathrm{H}}} \tag{9}$$

$$\nu = \frac{3B_{\mathrm{H}} - 2G_{\mathrm{H}}}{2(3B_{\mathrm{H}} + G_{\mathrm{H}})} \tag{10}$$

The Hill bulk moduli decrease from 364.54 GPa for $TcB_2$ to 328.32 GPa for $Tc_2B$ and 315.08 GPa for $Tc_3B$ . In contrast, shear and Young's moduli follow $TcB_2$ > $Tc_3B$ > $Tc_2B$. $Tc_3B$ thus resists shear deformation more strongly than $Tc_2B$ despite its smaller bulk modulus, showing that volumetric and shape-changing stiffness do not follow a single composition dependent ranking. Table 3 summarizes bulk mechanical descriptors.

**Table 3.** Voigt-Reuss-Hill moduli, empirical ductility indicators and Chen/Tian model hardness (GPa where applicable), with corresponding-pressure data reported by Tao et al. [28].

| Comp. | $B_V$ | $B_R$ | $B_H$ | $G_V$ | $G_R$ | $G_H$ | $E$ | $\nu$ | $B/G$ | $H_V^{\mathrm{Chen}}$ | $H_V^{\mathrm{Tian}}$ | Ref. |
|---|---|---|---|---|---|---|---|---|---|---|---|---|
| $TcB_2$ | 364.94 | 364.14 | 364.54 | 172.34 | 150.36 | 161.35 | 421.81 | 0.31 | 2.26 | 12.08 | 13.32 | This work |
| $Tc_2B$ | 328.68 | 327.96 | 328.32 | 133.58 | 125.77 | 129.67 | 343.76 | 0.33 | 2.53 | 8.61 | 10.02 | |
| $Tc_3B$ | 315.13 | 315.03 | 315.08 | 150.61 | 140.01 | 145.31 | 377.85 | 0.30 | 2.17 | 11.88 | 12.96 | |
| $TcB_2$ | - | - | 362.0 | - | - | 151.9 | 399.8 | - | - | 10.6 | 12.0 | [28] |
| $Tc_2B$ | - | - | 318.8 | - | - | 146.5 | 381.1 | - | - | 11.8 | 13.0 | |

| Comp. | $B_V$ | $B_R$ | $B_H$ | $G_V$ | $G_R$ | $G_H$ | $E$ | $\nu$ | $B/G$ | $H_V^{\text{Chen}}$ | $H_V^{\text{Tian}}$ | Ref. |
|---|---|---|---|---|---|---|---|---|---|---|---|---|
| $Tc_3B$ | - | - | 311.7 | - | - | 133.8 | 351.1 | - | - | 10.0 | 11.3 | |

**Figure 2(a)** highlights the distinct bulk- and shear-stiffness rankings. The literature comparison **[28]** gives relatively close bulk moduli, while shear-dependent quantities show greater differences.

### 3.2.3 Ductility and theoretical hardness

The Pugh ratio B/G was considered as an empirical ductility/brittleness indicator **[48]**, rather than a direct measure of fracture behavior. All three values exceed the conventional 1.75 threshold, and $Tc_2B$ has the largest B/G (2.532) and Poisson's ratio (0.326). It therefore exhibits the greatest elasticity-based ductile tendency within this set [**Figure 2(b)**], while $Tc_3B$ has the lowest B/G (2.168).

The Chen hardness estimator was evaluated from the Hill moduli **[49]**:

$$H_V^{\text{Chen}} = 2(k^2 G)^{0.585} - 3, \qquad k = \frac{G}{B} \tag{11}$$

The corresponding Tian hardness model is **[50]**:

$$H_V^{\text{Tian}} = 0.92\, k^{1.137} G^{0.708} \tag{12}$$

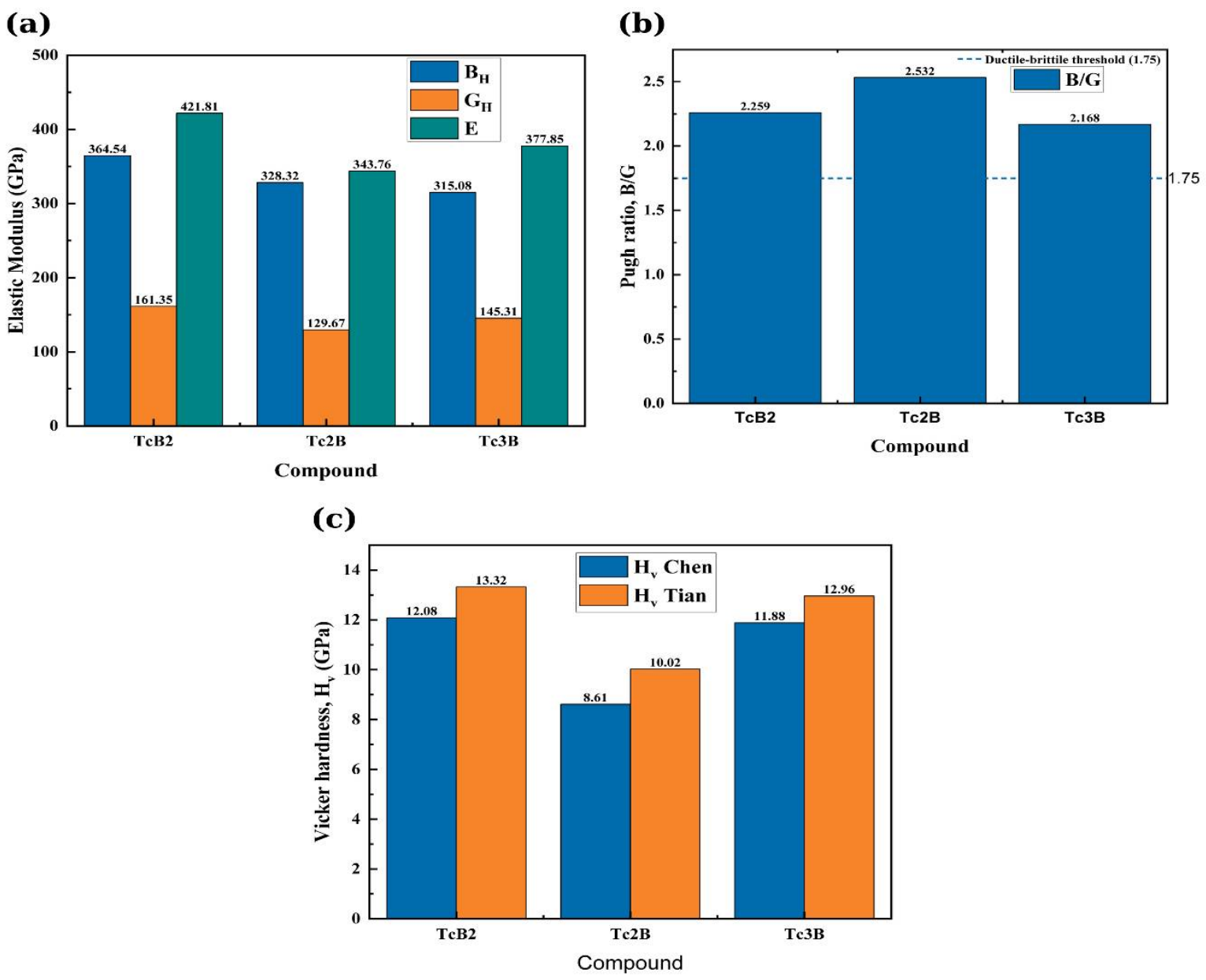

**Figure 2.** Comparative polycrystalline mechanical response of $TcB_2$ (20 GPa), $Tc_2B$ (0 GPa), and $Tc_3B$ (0 GPa): (a) Hill bulk modulus $B_H$, Hill shear modulus $G_H$, and Young's modulus $E$; (b) Pugh ratio $B/G$, with the empirical ductile–brittle boundary at 1.75; and (c) Vickers hardness $H_V$ estimated using the Chen and Tian models. The distinct 20 GPa condition of $TcB_2$ should be considered when separating pressure effects from stoichiometry- and symmetry-dependent trends.

Both models predict the same hardness order, $TcB_2 > Tc_3B > Tc_2B$. Their respective Chen/Tian estimates are 12.08/13.32, 11.88/12.96 and 8.61/10.02 GPa. **Figure 2**(c) illustrates this hierarchy and its association with shear resistance. These results are elasticity-based estimates, not indentation measurements and they do not support a superhard designation.

Taken together, **Figure 2** shows that the largest bulk modulus does not necessarily correspond to the greatest relative resistance to shear, while the largest elasticity-based ductility indicators accompany the smallest theoretical hardness in $Tc_2B$.

### 3.2.4 Elastic anisotropy and directional response

The universal elastic anisotropy index $A^U$ was evaluated following Ranganathan and Ostoja-Starzewski **[34]**:

$$A^U = 5\frac{G_V}{G_R} + \frac{B_V}{B_R} - 6 \tag{13}$$

An isotropic aggregate has $A^U = 0$. The calculated indices are 0.733, 0.313 and 0.379 for $TcB_2$, $Tc_2B$ and $Tc_3B$, respectively (**Table 4**), giving the ordering $TcB_2 > Tc_3B > Tc_2B$. For comparison, $A^U$ values reconstructed from the published single-crystal stiffnesses in Ref. **[28]** are listed separately and explicitly identified as derived values, not quantities directly tabulated in that study.

**Table 4.** Universal elastic anisotropy index $A^U$ for the investigated Tc-B phases and literature comparisons. Derived indices were recalculated from published $C_{ij}$ **[28]**; the separate ambient-pressure $AlB_2$-type $TcB_2$ value is from Ref. **[51]**.

| Compound | P (GPa) | $A^U$ | Reference |
|---|---|---|---|
| $TcB_2$ | 20 | 0.733 | This work |
| | | 1.237 | **[28]** |
| $TcB_2$ ($AlB_2$ type) | 0 | 3.047 | **[51]** |
| $Tc_2B$ | 0 | 0.313 | This work |
| | | 0.140 | **[28]** |
| $Tc_3B$ | 0 | 0.379 | This work |
| | | 0.245 | **[28]** |

The nonzero $A^U$ values establish elastic anisotropy in all three phases, with the largest overall index for $TcB_2$ and the smallest for $Tc_2B$. This scalar measure is complemented by the three-dimensional directional surfaces presented in **Figure 3**.

**Figure 3** depicts the directional dependence of Young's modulus E, shear modulus G, linear compressibility β and Poisson's ratio ν. The pronounced distortion of the Young's-modulus surface for $TcB_2$ and the lobed shear and Poisson-response surfaces for $Tc_3B$ highlight distinct directional deformation patterns. In contrast, the linear-compressibility surfaces are comparatively close to spherical. Thus, $A^U$ captures the overall departure from isotropy, whereas the surfaces reveal how the response varies with deformation mode.

Overall, $TcB_2$ is the stiffest of the investigated states at 20 GPa, $Tc_2B$ exhibits the strongest empirical ductile tendency, and $Tc_3B$ combines a smaller bulk modulus with a greater shear modulus and model hardness than $Tc_2B$ . These differences provide a compact mechanical context for the subsequent acoustic and thermal estimates.

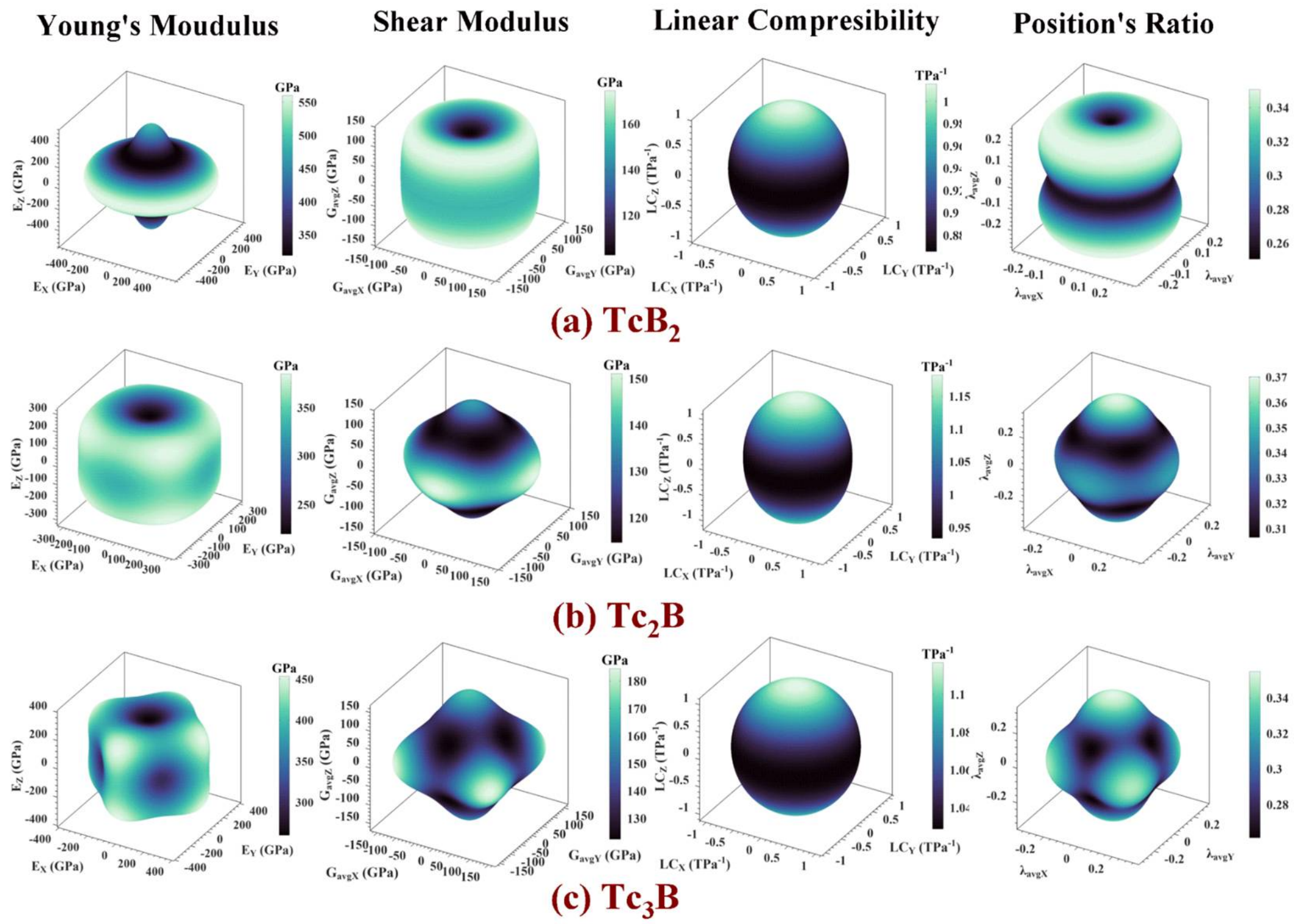

**Figure 3.** Three-dimensional directional dependence of Young's modulus $E$, shear modulus $G$, linear compressibility $\beta$, and Poisson's ratio $v$ for (a) $TcB_2$ at 20 GPa, (b) $Tc_2B$ at 0 GPa, and (c) $Tc_3B$ at 0 GPa. Deviations from a spherical shape indicate elastic anisotropy, while the color scales represent the magnitudes of the corresponding directional elastic properties.

### 3.2.5 Acoustic and Thermal Transport Properties

Elastic-derived acoustic and thermal parameters were evaluated using the optimized densities and Hill moduli. The 20 GPa condition of $TcB_2$ and the ambient-pressure conditions of $Tc_2B$ and $Tc_3B$ are maintained in all estimates. Related Ti-boride literature data are included in **Table 5** as contextual benchmarks, not as same-phase validations.

#### 3.2.5.1 Acoustic velocities and elastic Debye temperature

For an isotropic polycrystalline aggregate, the transverse and longitudinal velocities follow from the Hill moduli and density [35]:

$$v_t = \left(\frac{G_H}{\rho}\right)^{1/2} \tag{14}$$

$$v_l = \left( \frac{B_H + \frac{4}{3} G_H}{\rho} \right)^{1/2} \quad (15)$$

where ρ is the mass density, and $B_H$ and $G_H$ are the Hill moduli. The mean acoustic velocity is **[35]**:

$$v_m = \left[ \frac{1}{3} \left( \frac{2}{v_t^3} + \frac{1}{v_l^3} \right) \right]^{-1/3} \quad (16)$$

and the corresponding elastic Debye temperature, following Anderson **[35]**, is:

$$\Theta_D = \frac{h}{k_B} \left( \frac{3N}{4\pi V} \right)^{1/3} v_m \quad (17)$$

Here, h and $k_B$ are Planck's and Boltzmann's constants, respectively, and n and V are the number of atoms and volume of the selected cell. These quantities are continuum elastic estimates.

**Table 5.** Elastic-derived isotropic sound velocities and Debye temperatures, with published Ti-boride comparison values **[52]**. Velocities are in m $s^{-1}$; reference titanium borides are chemically and structurally different.

| Compound | $P$ (GPa) | $\rho$ (g $cm^{-3}$) | $v_l$ (m $s^{-1}$) | $v_t$ (m $s^{-1}$) | $v_m$ (m $s^{-1}$) | $\Theta_D$ (K) | Reference |
|---|---|---|---|---|---|---|---|
| $TcB_2$ | 20 | 8.165 | 8425.8 | 4445.4 | 4969.9 | 734.6 | This work |
| $Tc_2B$ | 0 | 10.167 | 7021.3 | 3571.3 | 4002.2 | 530.1 | |
| $Tc_3B$ | | 10.411 | 6991.0 | 3736.0 | 4173.0 | 538.8 | |
| $Ti_2B$ | | - | 8031.0 | 4552.0 | 5061.5 | 648.0 | [52] |
| $TiB_2$ | | - | 11585.9 | 7545.0 | 8272.4 | 1204.5 | |

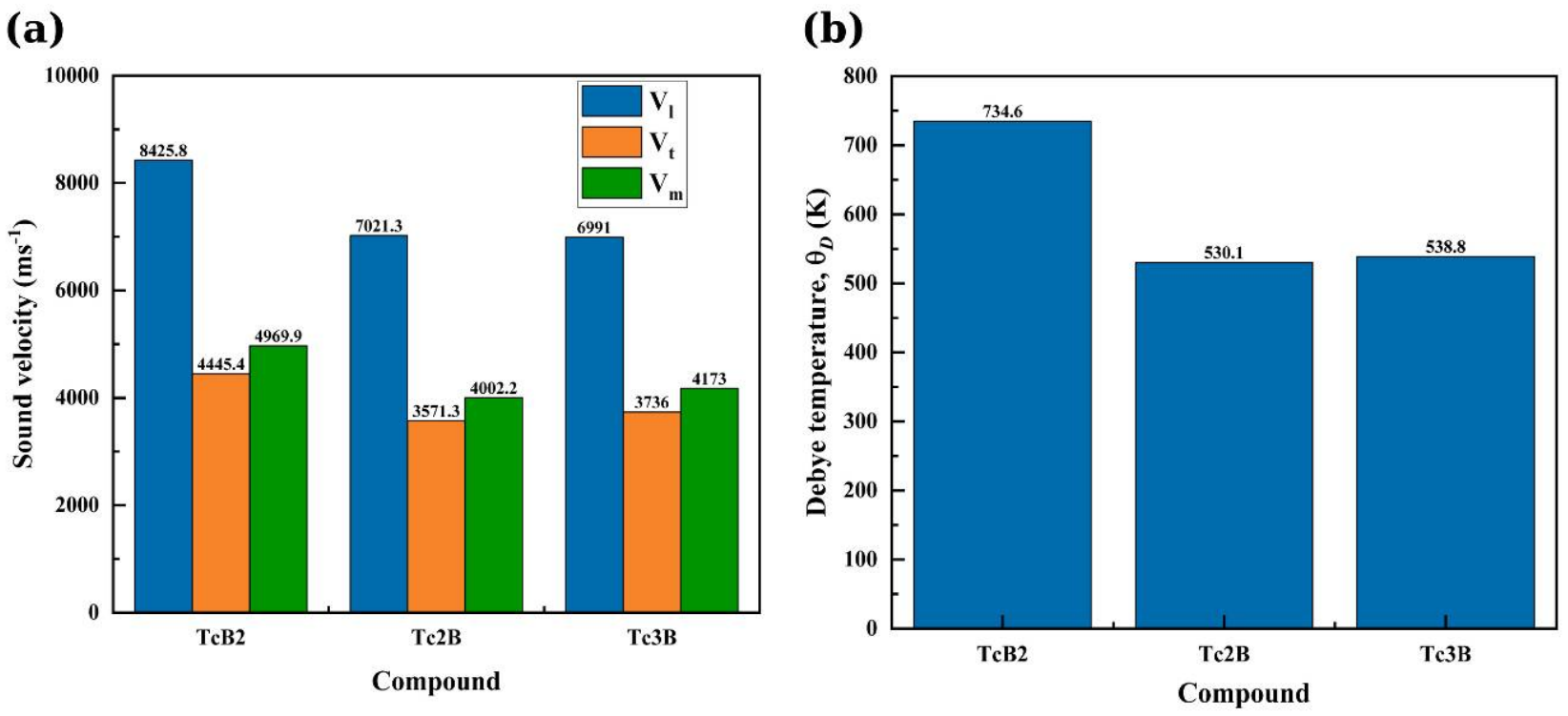

**Figure 4.** Elastic-derived acoustic response of $TcB_2$ (20 GPa), $Tc_2B$ (0 GPa), and $Tc_3B$ (0 GPa): (a) longitudinal, transverse, and mean acoustic velocities ($v_l$, $v_t$, and $v_m$) and (b) elastic Debye temperature $\Theta_D$. Pressure conditions are stated explicitly because $TcB_2$ is evaluated at 20 GPa, whereas $Tc_2B$ and $Tc_3B$ are evaluated at 0 GPa.

$TcB_2$ has the highest mean acoustic velocity (4969.9 m $s^{-1}$) and elastic Debye temperature (734.6 K) at 20 GPa. At ambient pressure, $Tc_3B$ has a slightly higher mean velocity and Debye temperature than $Tc_2B$, even though their longitudinal velocities are similar. The larger transverse contribution of $Tc_3B$ reflects its higher shear modulus.

**Figure 4** (a) compares the longitudinal, transverse and mean velocities, while **Figure 4** (b) shows the resulting Debye-temperature ranking. The parallel increase of $v_m$ and $\Theta_D$ emphasizes their common elastic origin; the higher $TcB_2$ values should nevertheless be interpreted in the context of its compressed state. The analogous Ti-boride data in **Table 5**, illustrate the sensitivity of these scales to composition and bonding, rather than providing direct Tc-phase benchmarks.

### 3.2.5.2 Directional acoustic response and minimum thermal conductivity

Principal-direction acoustic speeds were estimated from the single-crystal elastic coefficients and density. Along [100], the longitudinal and two transverse modes are governed by $C_{11}$, $C_{66}$ and $C_{44}$, respectively and along [001], they are governed by $C_{33}$ and the degenerate $C_{44}$ modes. A direction-specific, high-temperature adaptation of Cahill's minimum thermal conductivity model **[53]** was used:

$$\kappa_{\min}^{\text{dir}} = \frac{k_B}{2.48} n_a^{2/3} (v_L + v_{T1} + v_{T2}) \tag{18}$$

where $n_a$ is the atomic number density and $v^L$, $v^{T1}$ and $v^{T2}$ denote the three acoustic branches along the specified direction.

**Table 6.** Principal-direction acoustic velocities (km $s^{-1}$) and direction-specific Cahill-type minimum-conductivity estimates (W $m^{-1}$ $K^{-1}$). $TiB_2$ comparison values were recalculated using the elastic constants and structural parameters reported in Ref. **[52]**.

| Compound | $v_L^{[100]}$ | $v_{T1}^{[100]}$ | $v_{T2}^{[100]}$ | $v_L^{[001]}$ | $v_T^{[001]}$ | $\kappa_{\min}^{[100]}$ | $\kappa_{\min}^{[001]}$ | Reference |
|---|---|---|---|---|---|---|---|---|
| $TcB_2$ | 9.088 | 5.282 | 3.546 | 8.596 | 3.546 | 2.459 | 2.153 | This work |
| $Tc_2B$ | 7.085 | 4.044 | 3.674 | 6.424 | 3.674 | 1.631 | 1.517 | |
| $Tc_3B$ | 7.281 | 3.349 | 4.034 | 6.476 | 4.034 | 1.535 | 1.523 | |
| $TiB_2$ | 12.046 | 8.012 | 7.598 | 10.136 | 7.598 | 3.676 | 3.367 | **[52]** |

The [100] and [001] lower-limit values differ most in $TcB_2$ (2.459 versus 2.153 W $m^{-1}$ $K^{-1}$), less in $Tc_2B$ (1.631 versus 1.517 W $m^{-1}$ $K^{-1}$), and only marginally in $Tc_3B$ (1.535 versus 1.523 W $m^{-1}$ $K^{-1}$). These direction-specific comparisons concern acoustic-mode sums and do not replace the universal elastic-anisotropy assessment above.

### 3.2.5.3 Anharmonicity and lattice thermal conductivity

An acoustic Grüneisen parameter was estimated from Poisson's ratio using the quasi-isotropic Belomestnykh–Tesleva relation **[54]**:

$$\gamma = \frac{3(1+\nu)}{2(2-3\nu)} \tag{19}$$

The resulting values are $\gamma$ = 1.818, 1.943 and 1.773 for $TcB_2$, $Tc_2B$ and $Tc_3B$. As model-derived indicators, they imply the greatest effective anharmonic tendency in $Tc_2B$.

The minimum thermal conductivity was estimated using Clarke's model **[55]**:

$$\kappa_{\min}^{\text{Clarke}} = 0.87 k_B M_a^{-2/3} E^{1/2} \rho^{1/6} \tag{20}$$

where $M_a$ is the mean atomic mass and E is Young's modulus. The isotropic Cahill model [53] gives,

$$\kappa_{\min}^{\text{Cahill}} = \frac{k_B}{2.48} n_a^{2/3} (v_l + 2v_t) \tag{21}$$

The lattice thermal conductivity at 300 K was additionally estimated with the Slack model [36,56]:

$$\kappa_L = A \frac{\overline{M} \Theta_D^3 \delta}{\gamma^2 N_p^{2/3} T} \tag{22}$$

Here $\overline{M}$ is the average atomic mass (amu atom$^{-1}$), δ is the cubic root of atomic volume (Å), $N_p$ is the primitive-cell atom count, T is the temperature and A = 3.1 × $10^{-6}$ under the adopted units. The $\Theta_D^3$ and $\gamma^{-2}$ dependences link the estimate to lattice stiffness and effective anharmonicity.

At 300 K, the Slack estimates are 48.21, 19.31 and 36.16 W m$^{-1}$ K$^{-1}$ for $TcB_2$, $Tc_2B$ and $Tc_3B$. Their ordering is consistent with the combined effects of the elastic Debye temperature and effective Grüneisen parameter. Clarke and Cahill describe limiting estimates, whereas Slack approximates a lattice contribution and does not include electronic heat transport.

**Table 7.** Elastic-based anharmonic and thermal-transport estimates at 300 K (conductivities in W m$^{-1}$ K$^{-1}$), with $TiB_2$ minimum-conductivity data from the JARVIS-DFT record JVASP-20096 (database overview [57]).

| Compound | *P* (GPa) | $\gamma$ | $\kappa_{\min}^{\text{Clarke}}$ | $\kappa_{\min}^{\text{Cahill}}$ | $\kappa_L^{\text{Slack}}$ (300 K) | Reference |
|---|---|---|---|---|---|---|
| $TcB_2$ | 20 | 1.818 | 2.13 | 2.38 | 48.21 | This work |
| $Tc_2B$ | 0 | 1.943 | 1.38 | 1.56 | 19.31 | This work |
| $Tc_3B$ | 0 | 1.773 | 1.36 | 1.51 | 36.16 | This work |
| $TiB_2$ | 0 | - | 3.22 | 3.51 | - | [57] |

The $TiB_2$ database values in **Table 7** are separate, method-dependent minimum-conductivity estimates; they are not measured or directly interchangeable with the 300 K Slack values. **Figure 5** (a) displays the Clarke/Cahill lower limits, whereas **Figure 5** (b) compares the Slack estimates. $Tc_2B$ has the lowest model lattice conductivity despite having a higher bulk modulus than $Tc_3B$.

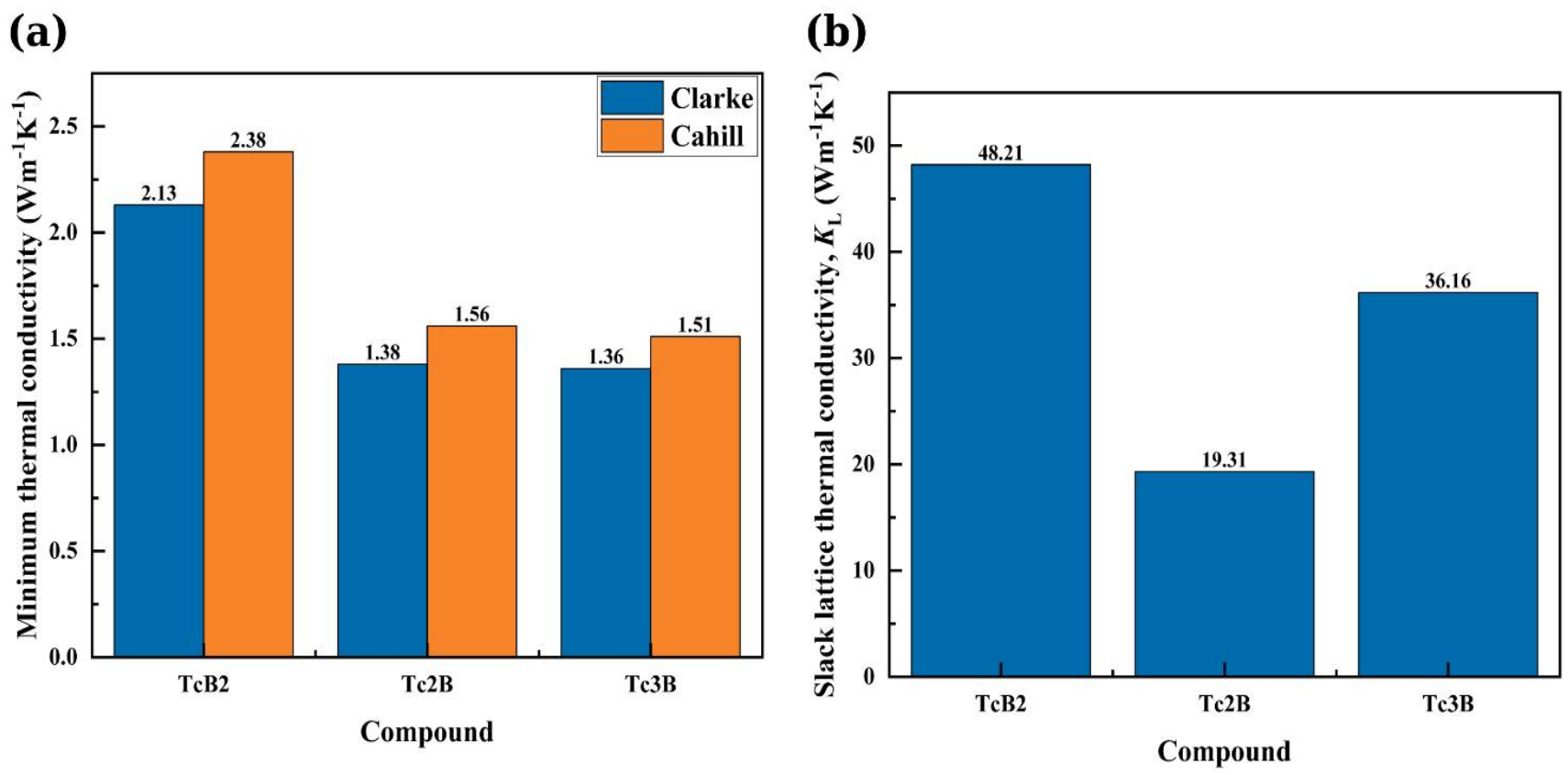

**Figure 5.** Elastic-derived thermal-transport descriptors of $TcB_2$ (20 GPa), $Tc_2B$ (0 GPa), and $Tc_3B$ (0 GPa): (a) minimum thermal conductivity $k_{min}$ estimated using the Clarke and Cahill models and (b) Slack lattice thermal conductivity $\kappa_L$ at 300 K. The Clarke/Cahill quantities represent lower-limit estimates, whereas the Slack value is a model-based lattice contribution rather than measured total thermal conductivity.

#### 3.2.6 Comparative thermal-transport assessment

The acoustic and thermal descriptors reinforce the shear-stiffness differences without implying a direct superconducting ranking. At the common ambient-pressure condition, $Tc_3B$ has higher transverse and mean sound velocities, elastic Debye temperature, and Slack-model conductivity than $Tc_2B$. $TcB_2$ exhibits the largest values under compression. Since these compounds are metallic **[28]**, total thermal transport should contain an electronic contribution, particularly at low temperatures. The present lattice estimates should not be regarded as full phonon Boltzmann-transport solutions. The elastic Debye scale is also distinct from the logarithmic phonon frequency $\omega_{log}$ employed in the subsequent electron-phonon-coupling analysis.

## 3.3 Electronic Structure

### 3.3.1 Band Structure and density of states

The SOC and non-SOC electronic band structures of $TcB_2$ (*P6/mmm* , 20 GPa), $Tc_2B$ (*I4/mcm*, 0 GPa), and $Tc_3B$ (*P4/mmm*, 0 GPa) are compared in **Figure 6**. The band structure paths are Γ-A-H-K-Γ-M-L-H for $TcB_2$ and Z-A-M-Γ-Z-R-X-Γ for $Tc_2B$ and $Tc_3B$, with $E_F$ shifted to 0 eV. All three phases remain metallic after SOC is included. Along the plotted high-symmetry paths, four SOC Kramers-pair branches cross $E_F$ in $TcB_2$, three in $Tc_2B$, and six in $Tc_3B$ and finite DOS at $E_F$ independently supports this multiband metallicity. In a spin–orbit-coupled electronic band structure, Kramers pair branches signify two electronic states that are related by time-reversal symmetry and are degenerate at specific points in momentum space. They are manifestations of a fundamental quantum-mechanical symmetry and have important implications for spin physics, band topology, and topological materials. The adjacent DOS/PDOS panels support the band discussion. Near $E_F$, the states are dominated mainly by Tc-4*d* character, consistent with previous Tc-boride and related transition-metal-diboride studies **[17,20,22,28,58–60]** and distinct from the B-*p* derived σ-band physics of $MgB_2$ **[7,8,61]**. The agreement between the band crossings and the finite density of states (DOS) provides a consistent and detailed picture of the electronic structure near the Fermi level. The electronic band dispersions near the Fermi level differ markedly among the compounds. $TcB_2$ exhibits the most strongly dispersive band manifold, characterized by an A-centered electron-like branch, an H-K hole-like branch, and an additional electron-like minimum along Γ-M near the Fermi level ($E_F$). $Tc_2B$ contains a denser sequence of extrema, including M-centered electron-like and M-Γ hole-like dispersions, whereas $Tc_3B$ has the most crowded low-energy manifold, with several electron-like branches along Z-A and M-Γ together with an X-centered hole-like feature. These assignments refer only to curvature along the plotted directions and complete pocket topology is treated in the Fermi-surface section. The directional variation remains compatible with the three-dimensional dispersion reported for technetium borides **[28]**. Because $TcB_2$ is evaluated at 20 GPa whereas $Tc_2B$ and $Tc_3B$ are at 0 GPa, the contrast reflects the combined effects of stoichiometry, crystal symmetry, bonding, and pressure rather than pressure alone. The progressively richer low-energy structure also anticipates increasing Fermi-surface complexity.

SOC mainly reorganizes local degeneracies and near-crossings rather than changing the metallic ground state. In these nonmagnetic centrosymmetric systems, combined inversion and time-reversal symmetry (*PT*) preserves twofold spin degeneracy at each *k*, so the observed SOC effects correspond to reconstruction of different band pairs, orbital degeneracies or avoided crossings rather than generic spin splitting. Representative SOC-induced reconstructions are approximately 0.133 eV in $TcB_2$ on H-K, 0.141 eV in $Tc_2B$ at Γ, and 0.227 eV in $Tc_3B$ on Γ-Z. These local changes preserve the metallic crossings, with $Tc_3B$ showing the strongest SOC response in the studied low-energy window and SOC therefore modifies the fine structure much more than the overall metallic character.

$Tc_2B$ also hosts distinctive special features. At R and X, two *PT*-degenerate SOC band pairs coincide, with a representative fourfold level at +0.5123 eV and the approximately linear separation of neighboring branches makes these SOC-stable fourfold Dirac-like candidates, consistent with crystalline-symmetry-protected fourfold degeneracies in three-dimensional materials **[62]**.

No comparably robust SOC-stable fourfold feature is resolved for $TcB_2$ or $Tc_3B$ within the plotted ±2 eV window. The present results therefore identify Dirac-like and van-Hove-related candidates without establishing a definitive topological classification or a true van Hove singularity which requires detailed analysis of the $Z_2$ invariance beyond the scope of the present study.

Overall, the evolution from $TcB_2$ to $Tc_2B$ to $Tc_3B$ is non-monotonic. $TcB_2$ at 20 GPa retains a clean, strongly dispersive metallic manifold. $Tc_2B$ is distinguished by R/X fourfold Dirac-like candidates and the clearest extended flat-band feature and $Tc_3B$ has the richest Fermi-level manifold and strongest local SOC reconstruction.

Coexisting electron-like and hole-like crossings anticipate a multi-sheet Fermi surface, while finite Tc-derived states at $E_F$ provide phase space for electron-phonon scattering. The bands therefore establish the electronic basis for later Fermi-surface, optical, and EPC analyses, but cannot by themselves determine the coupling strength or $T_c$, which depend on the phonon spectrum and Eliashberg coupling function **[17,28,46]**.

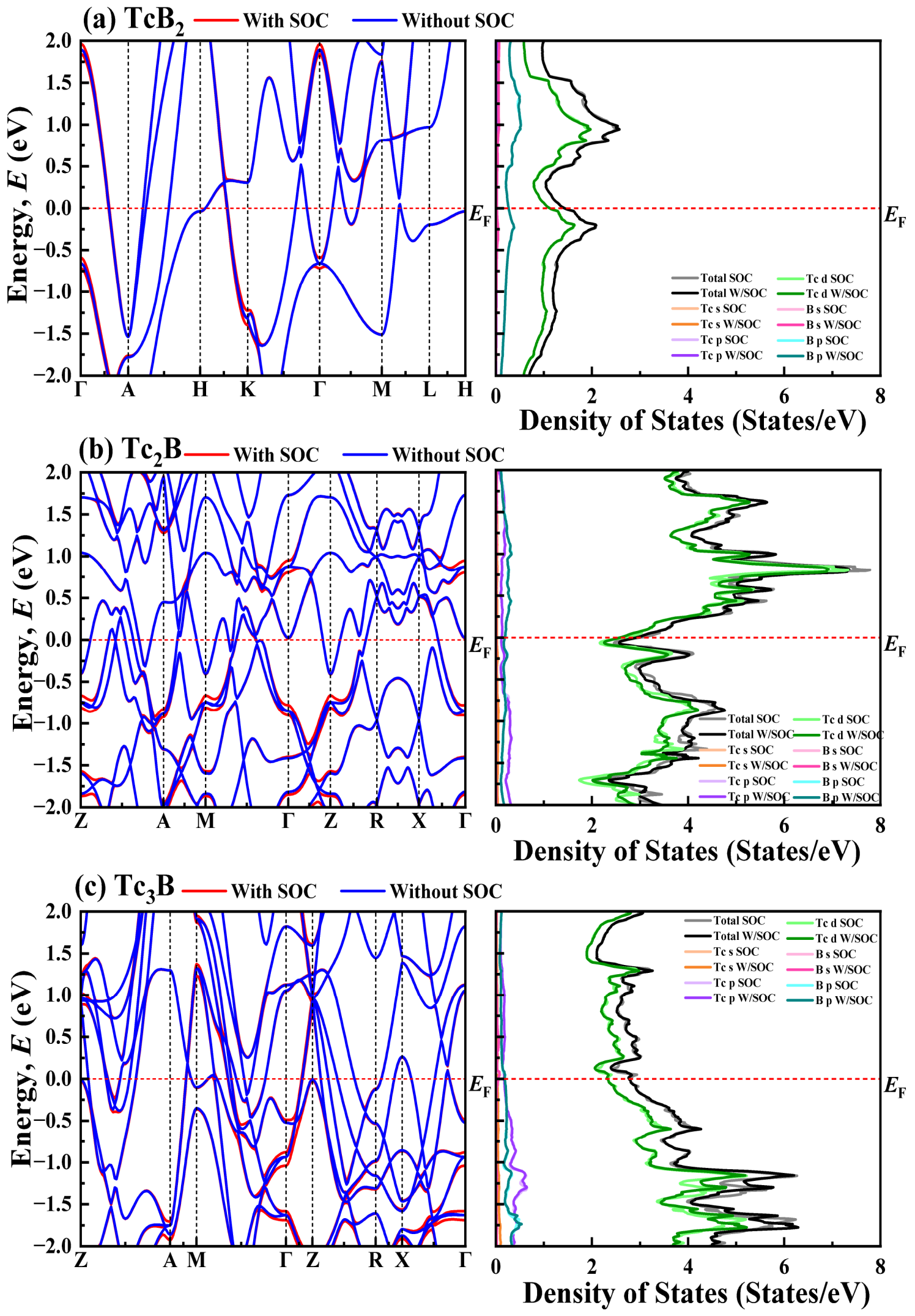

**Figure 6.** Electronic band structures (left) and corresponding total/projected density of states (right) of (a) $TcB_2$ at 20 GPa, (b) $Tc_2B$ at 0 GPa, and (c) $Tc_3B$ at 0 GPa. SOC bands are shown by red solid curves and non-SOC bands by blue solid curves; $E_F$ is set to 0 eV in all panels. The right-hand panels represents the DOS/PDOS.

### 3.3.2 Fermi Surface

**Figure 7** shows the Fermi-surface topologies of *P6/mmm* $TcB_2$ at 20 GPa, *I4/mcm* $Tc_2B$ at 0 GPa, and *P4/mmm* $Tc_3B$ at 0 GPa. Four bands (with band indices 10–13) cross the Fermi level in $TcB_2$, three (33–35) in $Tc_2B$, and six (21–26) in $Tc_3B$, consistent with multiband metallicity **[28]**.

The $TcB_2$ surface is strongly three-dimensional and warped, with band 11 forming the dominant extended sheet and bands 12 and 13 producing smaller closed pockets. Unlike $MgB_2$, it lacks weakly warped quasi-two-dimensional σ-band cylinders [7]. $Tc_2B$ has a simpler topology, combining extended sheets with compact pockets, whereas $Tc_3B$ shows the most intricate arrangement, with a large, warped band-25 sheet and several disconnected pockets. The nonuniform Fermi-velocity distribution suggests band-dependent anisotropy in charge transport. Since $TcB_2$ is considered at 20 GPa and the other phases at ambient pressure, these differences cannot be attributed solely to stoichiometry. Multiple Fermi-level sheets provide electronic states for electron-phonon scattering, but their number alone does not determine the coupling strength or superconducting transition temperature [60].

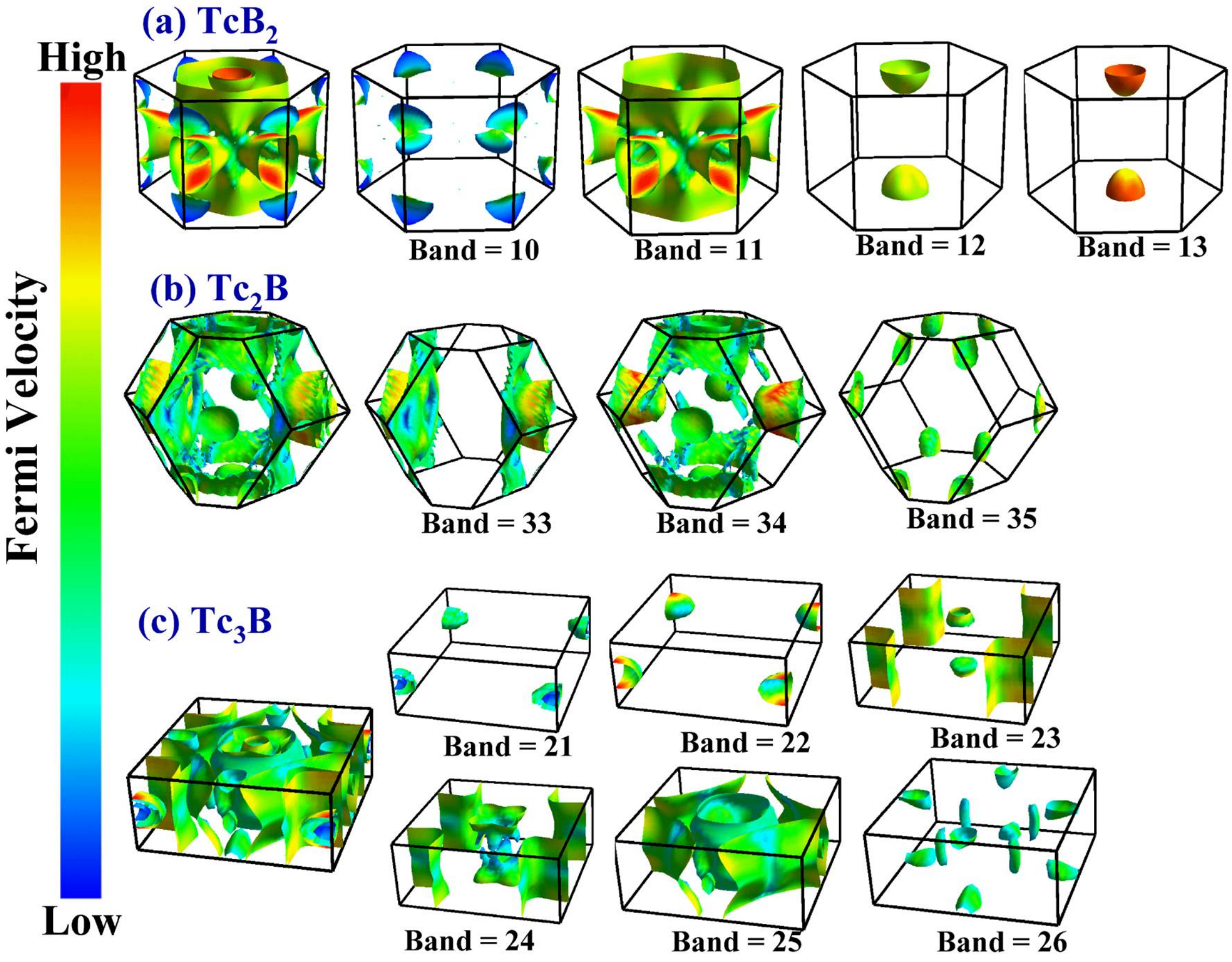

**Figure 7.** Band-resolved Fermi surfaces of (a) $TcB_2$ at 20 GPa, (b) $Tc_2B$ at 0 GPa, and (c) $Tc_3B$ at 0 GPa, calculated without SOC. Each row shows the combined surface and individual contributing bands. Color indicates relative Fermi-velocity magnitude within each dataset.

### 3.3.3 Charge Density

Figure 8 shows common-scale electron-density-difference (EDD) maps to compare the redistribution of valence charge upon bonding in $TcB_2$, $Tc_2B$, and $Tc_3B$. In $TcB_2$, the strongest positive redistribution of electronic charge is concentrated in the B-containing region, where distinctly directional accumulation between and around B sites is accompanied by depletion at neighboring Tc-centred regions. This pattern is consistent with substantial directional Tc-B/B-B bonding together with a partial charge-transfer component **[22,27,28,59]**. In the representative Tc-B-Tc plane of $Tc_2B$, accumulation remains associated with the B-containing environment, but it is less continuously developed, and the surrounding redistribution is more diffuse than in $TcB_2$. This supports a mixed picture of directional Tc- *d* /B-*p* covalent bonding with a larger metallic Tc-derived contribution **[22,63]**. $Tc_3B$ shows the clearest shift toward delocalization: Tc-centred depletion remains prominent, whereas compensating positive redistribution extends more broadly through interstitial regions of the Tc-rich network, compatible with greater delocalized Tc-derived electronic connectivity. This real-space trend is consistent with the DOS/PDOS picture, in which the low-energy states are predominantly Tc-4*d* derived while B-2*p* participation is relatively more conspicuous in B-rich $TcB_2$ **[22,28,59]**.

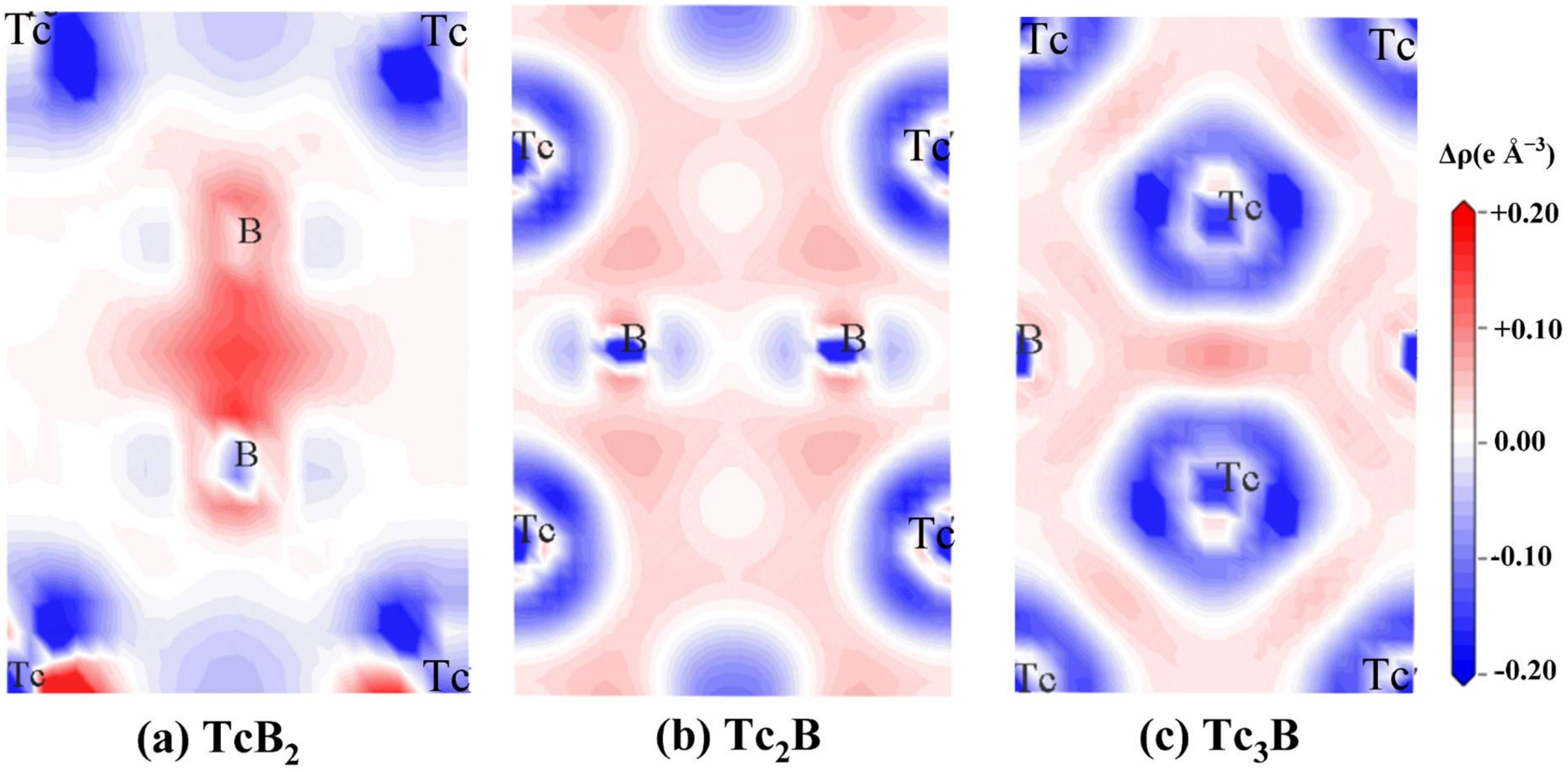

Figure 8. Electron density difference maps of $TcB_2$ [$(11\bar{2}0)$], $Tc_2B$ [representative Tc-B-Tc bonding plane], and $Tc_3B$ [(110)]. Red and blue indicate electron accumulation and depletion, respectively, with Δρ displayed over −0.20 to +0.20 e $Å^{-3}$.

Across $TcB_2$ → $Tc_2B$ → $Tc_3B$, the EDD maps therefore support a change in bonding emphasis from localized, directional B-associated redistribution toward a more spatially delocalized Tc-derived network, without reducing any phase to a purely ionic, covalent, or metallic limit. $TcB_2$ is evaluated at 20 GPa, whereas $Tc_2B$ and $Tc_3B$ are evaluated at 0 GPa; differences involving $TcB_2$ should therefore not be attributed solely to stoichiometry **[22,28]**. Qualitatively, the stronger directional B-associated redistribution is consistent with a more rigid bonding framework, whereas broader Tc-derived delocalization is consistent with a more metallic electronic environment **[27,64]**. These contrasting local bonding environments also form part of the electronic and force-constant background relevant to elasticity, lattice dynamics, and electron-phonon coupling.

### 3.4 Optical Properties

All three borides exhibit a pronounced metallic optical response at low photon energy, consistent with their finite Fermi-level electronic states and the dominant Tc-4d character around $E_F$ **[28]**. As summarized in **Figure 9**, the large low-energy imaginary part of the dielectric constant $\varepsilon_2(\omega)$, together with strong energy dependent structures in the real part of the dielectric constant $\varepsilon_1(\omega)$, refractive index n(ω), and extinction coefficient k(ω), are characteristics of metallic screening rather than an optical-gap onset **[65,66]**. Finite absorption and optical conductivity from the low-energy region into the visible range further support a combined free-carrier/interband response. At higher photon energies the spectra become increasingly structured: the lower-energy optical weight is consistent with transitions involving Tc-*d* dominated states near $E_F$, whereas higher-energy structures can include increasing contributions from B-*p*/Tc-*d* hybridized states.

Broad absorption and conductivity features extend from the visible into the ultraviolet, with the strongest absorption occurring mainly in the UV/deep-UV region. The comparative ordering is clear: $Tc_2B$ concentrates its principal absorption at lower photon energy, $Tc_2B$ shifts it to higher energy, and $Tc_3B$ lies between them. For [100] polarization of the electric field, representative maxima occur at 13.56 eV for $Tc_2B$ and 20.12 eV for $Tc_2B$, while $Tc_3B$ remains intermediate; the [001] spectra preserve the same broad compound-dependent separation. All three phases also retain substantial reflectivity across the visible range, consistent with their metallic electronic structures. The [100] and [001] curves separate in several spectral intervals in absorption, reflectivity, dielectric response, and optical conductivity, indicating moderate energy-dependent optical anisotropy expected for these non-cubic lattices and consistent with trends reported for related metallic borides **[67,68]**. Because $Tc_2B$ is evaluated at 20 GPa whereas $Tc_2B$ and $Tc_3B$ are evaluated at 0 GPa, its spectral differences cannot be attributed to stoichiometry alone.

A pronounced collective feature is common to the energy-loss spectra of all three compounds near 28 eV. The loss maxima coincide with high-energy zero crossings of $\varepsilon_1(\omega)$ while $\varepsilon_2(\omega)$ is reduced, supporting their interpretation as bulk-plasmon-like resonances rather than ordinary interband maxima **[65,66]**. The three phases therefore share a closely grouped collective-response energy scale, although the peak intensity and polarization dependence remain compound specific; $Tc_2B$ shows the largest loss intensity. This convergence of the loss features contrasts with the more widely separated absorption maxima and shows that the dominant interband spectral weight is more sensitive to composition, structure, polarization, and pressure condition than the high-energy collective resonance.

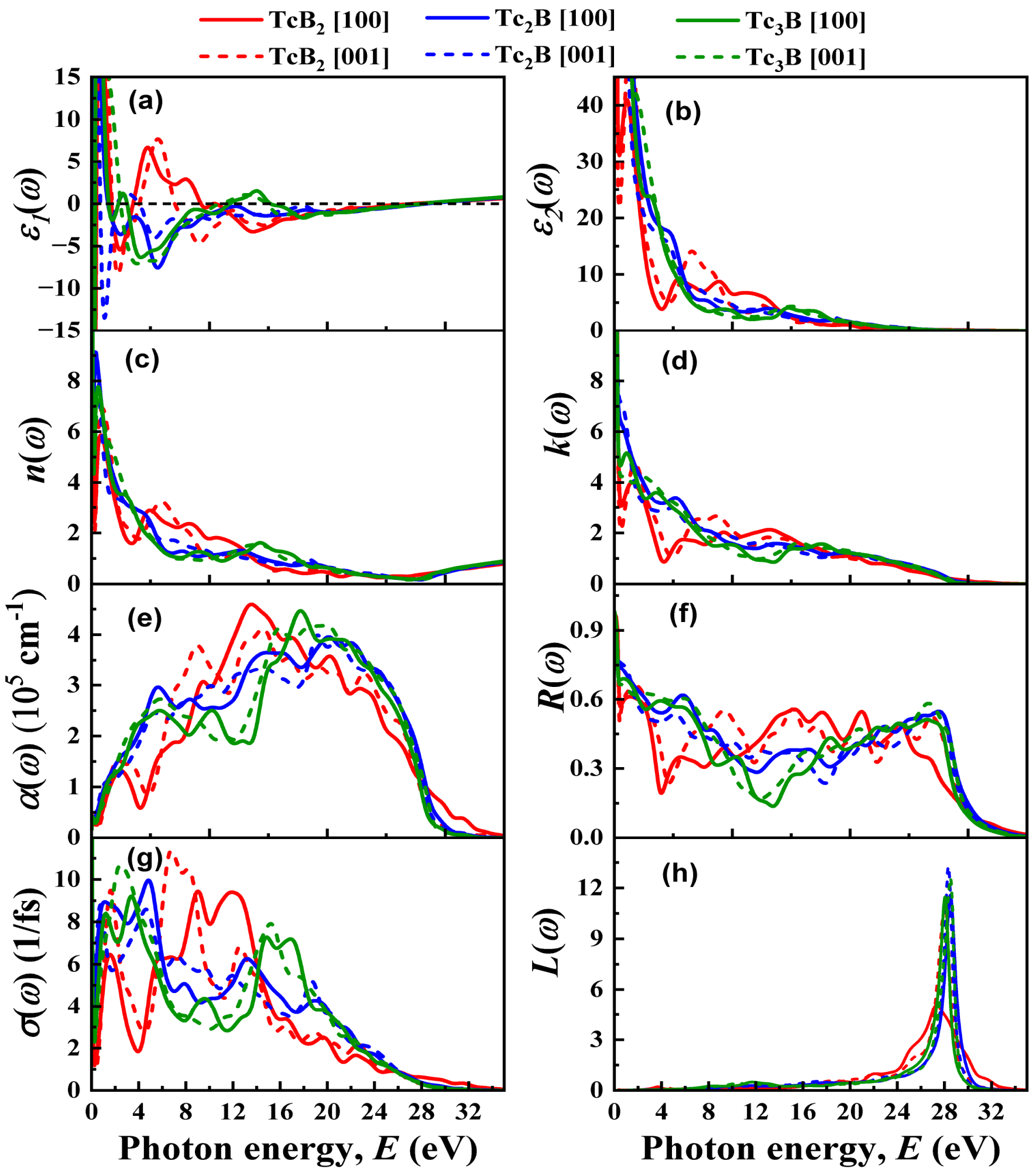

**Figure 9.** Calculated electric field polarization-dependent optical spectra of $TcB_2$ (*P6/mmm*, 20 GPa), $Tc_2B$ (*I4/mcm*, 0 GPa), and $Tc_3B$ (*P4/mmm*, 0 GPa) for [100] (solid lines) and [001] (dashed lines): (a) real dielectric function $\varepsilon_1(\omega)$, (b) imaginary dielectric function $\varepsilon_2(\omega)$, (c) refractive index $n(\omega)$, (d) extinction coefficient $k(\omega)$, (e) absorption coefficient $\alpha(\omega)$, (f) reflectivity $R(\omega)$, (g) optical conductivity $\sigma(\omega)$, and (h) electron energy-loss function $L(\omega)$.

## 3.5 Electron-Phonon Coupling and Superconductivity

**Figure 10** compares the direction-resolved phonon dispersions within the Brillouin zone and projected phonon density of states (PhDOS) with the corresponding EPC-weighted phonon spectra, Eliashberg spectral function $\alpha^2F(\omega)$, and cumulative coupling $\lambda(\omega)$ for $TcB_2$ at 20 GPa and for $Tc_2B$ and $Tc_3B$ at 0 GPa. No imaginary branches are observed, and the acoustic modes approach zero at Γ, confirming harmonic dynamical stability under the investigated conditions. The projected spectra place most Tc (Technetium) character in the low-frequency region, whereas B contributions become increasingly important at higher frequencies, particularly in $TcB_2$ and $Tc_3B$.

The EPC-resolved spectra show that the dominant coupling is concentrated mainly in the Tc-rich low-frequency sector. For $TcB_2$, the principal $\alpha^2F(\omega)$ feature occurs near 94 $cm^{-1}$ and the largest sampled linewidth enhancement is near 38 $cm^{-1}$; about 91% of the cumulative $\lambda(\omega)$ is accumulated below 180 $cm^{-1}$. $Tc_2B$ exhibits a broader EPC-active region over approximately 140–260 $cm^{-1}$, with structures near 173, 201, 236, and 256 $cm^{-1}$ and an additional high-frequency feature near 508 $cm^{-1}$; its largest sampled linewidth is about 64 $cm^{-1}$, and nearly 87% of the coupling is accumulated below 275 $cm^{-1}$. In $Tc_3B$ , the main EPC weight lies between about 130 and 225 $cm^{-1}$, the largest sampled linewidth is near 88 $cm^{-1}$, and approximately 89% of the cumulative coupling is reached below 250 $cm^{-1}$. The B-rich feature around 410 $cm^{-1}$ provides only a modest additional contribution. These results consistently identify low-frequency Tc-rich vibrations as the principal EPC-active sector in all three compounds.

**Table 8**. Calculated superconducting parameters and comparison of transition temperatures with previous theoretical results.

| Compound | Pressure (GPa) | λ | $\omega_{log}$ (K) | $T_c$ (K) | Literature Tc (K) | Reference |
|---|---|---|---|---|---|---|
| $TcB_2$ | 20 | 2.4388 | 153.405 | 24.927 | 23.1 | This work / [28] |
| $Tc_2B$ | 0 | 0.7771 | 251.131 | 11.078 | 10.9 | This work / [28] |
| $Tc_3B$ | | 0.9175 | 249.799 | 15.128 | 12.9 | This work / [28] |

Within density-functional perturbation theory [67], the Eliashberg spectral function $\alpha^2F(\omega)$ identifies the phonon frequencies that couple most effectively to electronic states at the Fermi level [2,68]. The total electron–phonon coupling constant is obtained from [2,70]:

$$\lambda = 2\int_0^{\infty} \frac{\alpha^2F(\omega)}{\omega}\, d\omega, \tag{23}$$

The accumulated coupling up to a phonon frequency Ω is described by,

$$\lambda(\Omega) = 2\int_0^{\Omega} \frac{\alpha^2F(\omega)}{\omega}\, d\omega, \tag{24}$$

The 1/ω weighting emphasizes low-frequency modes only when they carry appreciable EPC spectral weight. The logarithmic average phonon frequency is defined as:

$$\omega_{\log} = \exp\left[\frac{2}{\lambda}\int_0^{\infty} \frac{\alpha^2F(\omega)}{\omega} \ln\omega\, d\omega\right], \tag{25}$$

The superconducting transition temperature is then estimated using the simplified Allen–Dynes-modified McMillan expression [45,46,72]:

$$T_c = \frac{\omega_{\log}}{1.20}\exp\left[-\frac{1.04(1+\lambda)}{\lambda-\mu^*(1+0.62\lambda)}\right]. \tag{26}$$

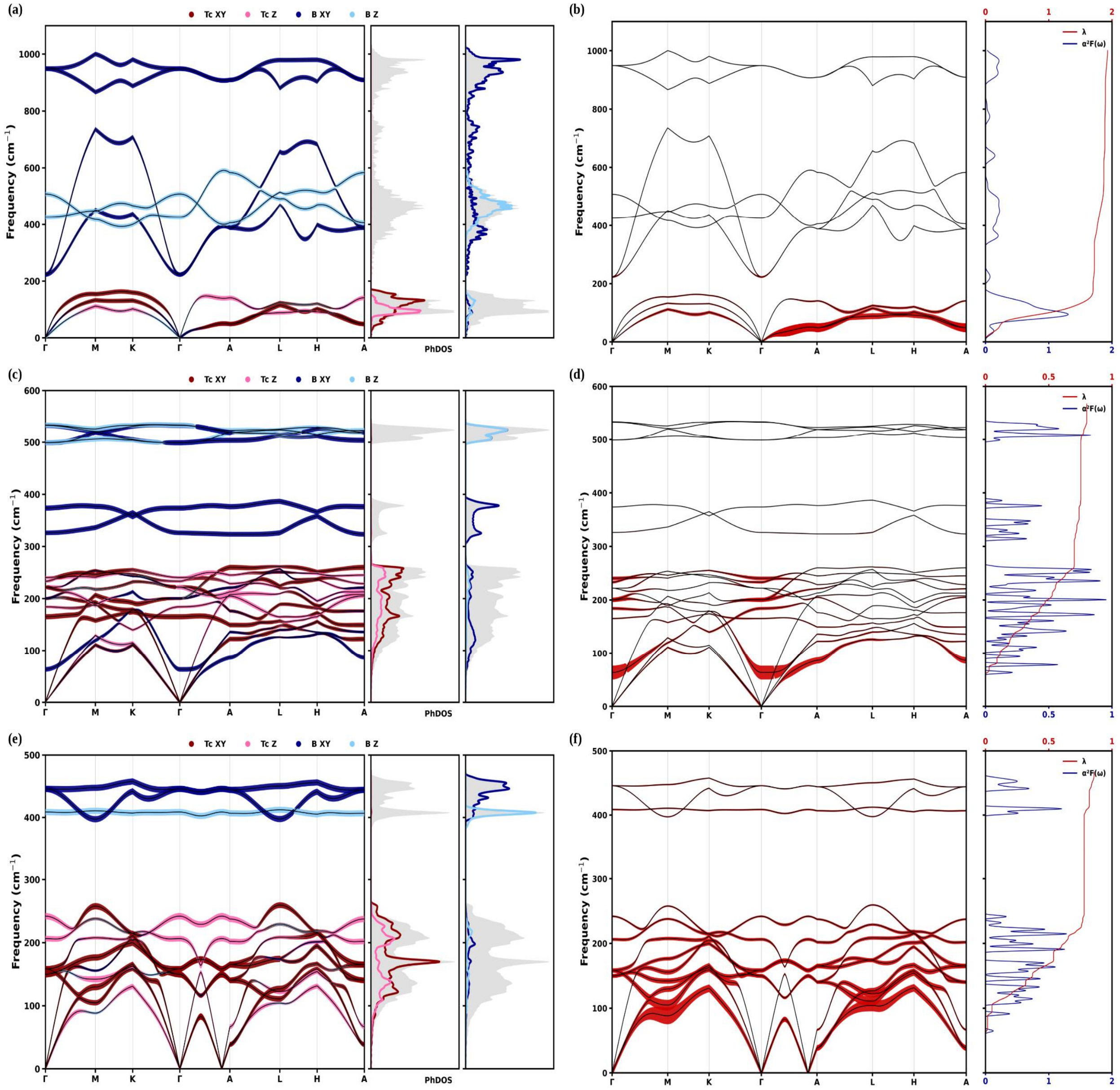

**Figure 10.** Phonon and electron–phonon coupling characteristics of $TcB_2$ at 20 GPa, $Tc_2B$ at 0 GPa, and $Tc_3B$ at 0 GPa. Panels (a), (c), and (e) show direction-resolved phonon dispersions and projected phonon density of states (PhDOS); panels (b), (d), and (f) show the corresponding EPC-weighted phonon dispersions together with the Eliashberg spectral function $\alpha^2F(\omega)$ and cumulative EPC $\lambda(\omega)$.

The calculated superconducting parameters show that $TcB_2$ shows the strongest coupling, $\lambda = 2.4388$, compared with 0.7771 for $Tc_2B$ and 0.9175 for $Tc_3B$. The logarithmic phonon frequencies follow the opposite trend, with $\omega_{log} = 153.405$ K for $TcB_2$, 251.131 K for $Tc_2B$, and 249.799 K for $Tc_3B$. Consequently, the large coupling in $TcB_2$ compensates for its lower characteristic phonon scale and gives the highest calculated transition temperature, 24.927 K. $Tc_3B$ reaches 15.128 K, above the 11.078 K of $Tc_2B$, because its larger λ outweighs their nearly identical $\omega_{log}$ values. The resulting $T_c$ hierarchy therefore reflects the balance among EPC strength, the EPC-weighted phonon scale, and Coulomb repulsion rather than any single phonon frequency **[2,45,46,72]**. The computed EPC of $TcB_2$ places this compound under the category of very strongly coupled superconductor class.

The present results shows that the transition-temperature ordering is $TcB_2$ > $Tc_3B$ > $Tc_2B$, while giving stronger coupling for $TcB_2$, slightly weaker coupling for $Tc_2B$, and close agreement for $Tc_3B$. Both studies nevertheless associate the dominant EPC with low-frequency Tc vibrations interacting with Tc-4*d* states near the Fermi level **[28]**.

The detailed superconducting mechanism of the three compounds differs from the canonical $MgB_2$ picture, where superconductivity is governed by strongly coupled B-derived *σ* states and B bond-stretching phonons **[11,73]**. In the present Tc borides, the electronic states near the Fermi level contain substantial Tc-derived weight and the dominant EPC is associated with Tc-rich low-frequency vibrations, a qualitative behavior closer to transition-metal borides such as pressurized $MoB_2$, where metal states and low-frequency metal vibrations are important **[17]**. Although the calculated Fermi surfaces are multiband, the isotropic $\alpha^2F(\omega)$ and total λ used here do not resolve band-specific pairing and therefore provide no basis for claiming multigap superconductivity.

Again, the cross-compound differences should not be attributed to stoichiometry alone because $TcB_2$ is evaluated at 20 GPa, whereas $Tc_2B$ and $Tc_3B$ are evaluated at 0 GPa, and no same-phase pressure series was calculated. The comparison therefore reflects the combined effects of Tc:B ratio, crystal symmetry, bonding topology, electronic states near the Fermi level, phonon bandwidth, and the distinct pressure condition of $TcB_2$. Within these limits, the calculations support conventional phonon-mediated superconductivity in all three phases, with Tc-rich low-frequency vibrations providing the principal EPC contribution and $TcB_2$ exhibiting the largest λ and highest predicted $T_c$.

The strongest transferable superconducting lesson comes from the correspondence between low-energy electronic states and EPC-active lattice vibrations. All three phases are multiband metals with predominantly Tc-4*d* states near $E_F$, yet neither $N(E_F)$ nor Fermi-surface complexity alone reproduces the superconducting hierarchy. $Tc_2B$ additionally illustrates the electronic diversity of the series through a Dirac-like candidate feature, without establishing a topological semimetal phase **[62]**. More generally, the stronger EPC response occurs when suitable Tc-derived electronic states couple effectively to dynamically stable, predominantly low-frequency Tc-rich phonons, emphasizing electronic-phonon matching rather than any single electronic descriptor **[2,17,28,59]**. The superconducting response therefore depends on the balance between coupling strength and the characteristic phonon energy, so maximizing either quantity independently is not a sufficient criterion **[2,49]**. $TcB_2$ is evaluated at 20 GPa, whereas $Tc_2B$ and $Tc_3B$ are evaluated at 0 GPa; differences involving $TcB_2$ consequently contain a pressure contribution in addition to composition, symmetry, and bonding and should not be interpreted as purely stoichiometric trends. Accordingly, the Tc-B series is most useful for identifying combinations of bonding environments, *d*-derived states, stable EPC-active phonons, and sufficiently strong coupling while maintaining dynamical stability that can be examined in closely related transition-metal borides.

# 4 Conclusions

This work establishes a consistent comparative picture of $TcB_2$ at 20 GPa and $Tc_2B$ and $Tc_3B$ at 0 GPa. The optimized structures closely reproduce the corresponding literature phases, and all three satisfy the relevant elastic criteria and are dynamically stable within the harmonic approximation under the investigated conditions. Within the scope of the present work, stability assessment is based on the elastic criteria and harmonic phonon calculations. $TcB_2$ is the most rigid member under compression, $Tc_2B$ shows the strongest empirical ductile tendency and the lowest model lattice-thermal estimate, and $Tc_3B$ occupies an intermediate mechanical regime. These contrasts accompany an electronic-bonding evolution from more directional B-associated charge redistribution in B-rich $TcB_2$ toward more delocalized Tc-derived connectivity in the Tc-rich phases. The compounds under consideration are fairly hard and highly machinable – a combination suitable for structural engineering [74-76] applications. All three are metallic, with low-energy states dominated by Tc-4d character and appreciable Tc-d/B-p hybridization. Their calculated optical responses are likewise metallic and electric field polarization dependent, with substantial reflectivity and strong ultraviolet response suitable for optical applications.

All three compounds are multiband metals, but their low-energy electronic structures differ substantially. $Tc_3B$ has the richest Fermi-surface topology and the strongest local SOC reconstruction in the studied energy window, whereas $Tc_2B$ exhibits SOC-stable fourfold Dirac-like candidates together with a flat-band/van-Hove-related feature. At the common 0.04 Ry comparison point, $TcB_2$ has $\lambda$ = 2.4388, $\omega_{log}$ = 153.405 K, and $T_c$ = 24.927 K; $Tc_2B$ has $\lambda$ = 0.7771, $\omega_{log}$ = 251.131 K, and $T_c$ = 11.078 K; and $Tc_3B$ has $\lambda$ = 0.9175, $\omega_{log}$ = 249.799 K, and $T_c$ = 15.128 K. Low-frequency Tc-rich vibrations provide the dominant EPC-active contribution. The superconducting response is better captured by electronic–phonon matching and the balance between $\lambda$ and $\omega_{log}$ than by any single electronic descriptor. Pressure can be used to further tune the superconducting state properties [77-79] of all the three compounds under consideration, particularly by pressure induced enhancement of the logarithmically averaged phonon frequency. This possibility should be explored in future studies.

Overall, the $TcB_2$-$Tc_2B$-$Tc_3B$ series shows that the principal property trends are non-monotonic and arise from the combined effects of composition, crystal symmetry, bonding topology, and pressure. In particular, comparisons involving $TcB_2$ must retain its distinct 20 GPa condition, whereas $Tc_2B$ and $Tc_3B$ are evaluated at 0 GPa. The broader value of this study lies in candidate structure–property guidelines for future boride screening and materials discovery. A useful design hypothesis is to search for transition-metal borides that reproduce favorable *d*-electron/EPC-active-phonon matching while retaining mechanical and dynamical stability and sufficiently strong coupling at substantially lower or ambient pressure. These criteria provide candidate screening directions for future validation and materials discovery.

## Data availability

The data sets generated and/or analyzed in this study are available from the corresponding author on reasonable request.

## Declaration of interest

The authors declare no competing interests.

## CRediT authorship contribution statement

**Md Rabi Us Sani Shapon:** Software, Methodology, Formal analysis, Data curation, Visualization, Writing-original draft, review & editing; **Jubair Hossan Abir:** Software, Formal analysis, Data curation, Visualization, Writing-draft, review & editing; **Md. Asif Afzal:** Formal analysis, Visualization, Writing-draft; **Raihana Shams Islam:** Conceptualization, Supervision, Validation, Administration, Writing- Reviewing and Editing; **Saleh Hasan Naqib:** Conceptualization, Supervision, Validation, Administration, Writing- Reviewing and Editing.

## Declaration of Generative AI and AI-assisted technologies in the writing process

During the preparation of this work, the authors used ChatGPT (OpenAI) to improve the language, grammar, readability, and organization of the manuscript. After using this tool, the authors reviewed and edited the content as needed and take full responsibility for the content of the published article.